\documentclass[twocolumn,aps,pra,10pt,eqsecnum,showpacs]{revtex4-1}
\usepackage{bm}
\usepackage{amsfonts}
\usepackage{amssymb}
\usepackage{amsmath}
\usepackage{graphicx}
\usepackage{epstopdf}
\begin{document}
\title{Limitations in the $2D$ description of the electromagnetic waves propagation in thin dielectric and magnetic layers}
\author{T. Radozycki}
\email{torado@fuw.edu.pl}
\affiliation{Center for Theoretical Physics, Polish Academy of Sciences, Al. Lotnik\'ow 32/46, 02-668 Warsaw, Poland}
\author{P. Bargiela}
\email{piotr.bargiela@gmail.com}
\affiliation{Center for Theoretical Physics, Polish Academy of Sciences, Al. Lotnik\'ow 32/46, 02-668 Warsaw, Poland}
\begin{abstract}
The propagation of electromagnetic waves trapped within dielectric and magnetic layers is considered. The description within the three-dimensional theory is compared with the simplified analysis in two dimensions. Two distinct media configurations with different topology are dealt with: a plane slab and a hollow cylinder. Choosing the appropriate values for the geometrical parameters (layer thickness, radius of the cylinder) and for the electromagnetic properties of the media one can trap exactly one mode corresponding to that obtained within the two-dimensional electromagnetism. However, the symmetry between electric and magnetic fields suggests, that the two versions of the simplified electromagnetism ought to be taken into account. Its usual form is incomplete to describe all modes. It is also found that there is a domain of optimal values of parameters for which the $2D$ model works relatively correctly. In the case of a cylindrical surface we observe, however, several differences which are attributed to the curvature of the layer, and which exclude the propagation of evanescent modes. The two-dimensional electrodynamics, whichever form used, turns out still too poor to describe the so called `hybrid modes' excited in a real layer.  The obtained results can be important for proper description of the propagating waves within thin layers for which $3D$ approach is not available due to mathematical complexity, and reducing the layer to a lower-dimensional structure seems the only possible option.
\end{abstract}
\maketitle
\section{Introduction}\label{intro}

Physics in thin layers (or thin films) has been increasingly gaining in importance due to the miniaturization in electronics, the emergence of new materials like graphene, the use of ultra-thin coatings in medicine. In particular the attention of physicists has been attracted by the propagation of light through dielectric layers of thickness comparable to the length of the waves~\cite{knittel} leading to the so called `thin layer optics'. The practical applications comprise the reflection of waves on the edges of a medium and interference leading to the appearance of anti-reflective coatings, thin lenses, narrow-band filters, beam splitters or high-reflectivity mirrors to mention only a few~\cite{raut,mcleod,ar,pf,pro}.

The question of correct description of phenomena emerging in thin layers is a longstanding one and refers to many branches of physics: classical and quantum mechanics, statistical physics, electrodynamics and so on~\cite{zang,des,luth}. One approach is the direct formulation of the lower-dimensional theory. The opposite alternative is to deal with the full three-dimensional theory and to apply in the final results a limiting procedure with respect to the contracted dimension. The classical and quantum mechanical calculations show, however, some difference in the reduced theory formulated directly in $2D$ space and that obtained as a limit of a shrinking width in $3D$~\cite{vankamp}. An intermediate approach consists in deriving fundamental equations governing the system and then obtaining their approximate form, which can be solved either analytically or numerically~\cite{ly1,ly2,yz}.

In the present paper we would like to concentrate on waves propagating inside the media, which may be considered as `trapped waves' due to the value of refractive index larger than $1$. Such a situation is often described within the two-dimensional theory of electromagnetism~\cite{hillion,lapidus,zwiebach} which is treated as a model of such a layer. It is commonly assumed that this kind of description is adequate to account for essential properties of waves trapped between the walls.

The goal of this paper is to analyze the propagating modes and to establish if, and under what conditions, a thin layer may be considered as a two-dimensional structure. We do not expect discrepancies of the kind observed in~\cite{vankamp} to appear, since they were connected with the infinite potential walls responsible for external constraints. There is no place for such an effect in a dielectric layer (which is thereby quite similar to a finite potential well), unless the refractive index $n$ is extremely large. This in turn would make the light velocity almost zero and it would extremely shorten the wavelength. We assume $n$ to be larger than one, either due to dielectric or magnetic properties of the medium, but would rather concentrate on wave frequency $\omega$, layer thickness $d$ and its curvature as free parameters. We also neglect any energy losses in the dielectric or magnetic medium.

Propagating waves considered here do not have any sources, or more precisely, their sources are very distant. Therefore, in the present paper we do not take into account charges, currents and their interactions. It is then not a complete $2D$  electrodynamics. If one liked to fully reproduce it in a layer, other phenomena should be considered as well. For instance the $3D$ Gauss law should be reduced to the world in 2D and electric field lines coming from charges should be trapped within the layer so that the resulting Coulomb force is described as $F\sim 1/r$ instead of the familiar $1/r^2$.

The waves mentioned above correspond to evanescent solutions of Maxwell equations, and do not appear as a result of an illumination of the layer from the outside. They must rather be generated inside. The external modes can freely propagate in both media, but they are not of any interest for us. They are a simple consequence of the fact that our system is immersed in a wider three-dimensional space.

The simplest system of that kind is a thin rectangular slab with dielectric and/or magnetic properties. It will be dealt with in Sec.~\ref{slab}. Even in this case we find rather firm restrictions that must be imposed on the system to be treated as a two-dimensional one. Moreover, depending on the properties of the media, an alternative $2D$ theory has to be used. In Sec.~\ref{cyl} we consider a bit more complicated structure, namely that of the thin cylindrical layer or, in other words, a hollow fiber. The results derived from the two-dimensional version of the theory are basically consistent with those obtained in 3D under similar requirements as found for the slab. However, there are some differences attributable to the surface topology and curvature. Not all the modes allowed on the surface have their counterparts in a cylindrical layer. And vice-versa, the so called `hybrid' modes propagating in real world do not find their two-dimensional representation.

Besides, for tightly bent cylinder, some waves have a tendency to escape from the layer, which, of course, cannot take place in the $2D$ world. Even the simplest mode propagating along the cylinder axis has a minimal value of frequency, below which this propagation is not possible. This limiting value emerges as a consequence of the surface curvature and depends on the cylinder radius; it completely disappears when cylindrical layer becomes flat, ($R\to\infty$). In the $2D$ model, no such threshold appears. The same effect leads to the modification of thresholds existing in $2D$ in the case of helical propagation (i.e. with azimuthal number $m$ different from zero).

Throughout the paper we use the system of units, where
$c=1$ and $\hbar=1$. In such a system the vacuum permeability is the inverse of permittivity:
$$
\mu_0=(\epsilon_0)^{-1}.
$$
Electric and magnetic fields are written in terms of complex functions, and the corresponding physical quantities are obtained by taking the real parts.

\section{Rectangular slab}\label{slab}

\subsection{Real propagation in a planar layer}\label{rl}

\begin{figure}[h!]
\begin{center}
\includegraphics[width= 0.45\textwidth,angle=0]{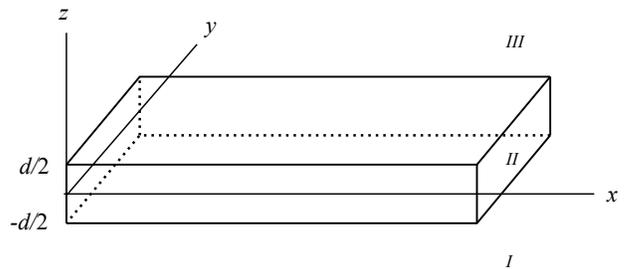}
\end{center}
\caption{Dielectric/magnetic slab and definitions of regions $I$, $II$, $III$.}
\label{figslab}
\end{figure}

Let us consider an infinite rectangular slab of thickness $d$, either dielectric (with relative permittivity $\epsilon>1$ and relative permeability $\mu= 1$), or magnetic (with $\epsilon=1$ and $\mu> 1$). The medium fills the region between the planes at $z=-d/2$ and $z=d/2$, and divides the whole space onto three regions $I$, $II$ and $III$ as shown in Fig.~\ref{figslab}.

When the medium has a large refractive index $n=\epsilon\mu$, some of the modes propagating inside cannot escape due to the total internal reflection. Such modes are referred to as `evanescent' modes and they decay exponentially with perpendicular distance in the space outside the slab. This feature discriminates them from the waves propagating in a conducting waveguide. Below we verify whether and under what conditions the wave captured within a slab corresponds to a two-dimensional propagation. Surely the slab is immersed in the $3D$ space, so there are plenty of (external or radiative) modes, passing through it. They do exist independently on what refractive index or thickness of the slab is chosen. They undergo refraction, but none of them can be captured inside, so they do not participate in modeling the $2D$ space.

All waves satisfy the sourceless Maxwell equations:
\begin{subequations}\label{max}
\begin{align}
&{\bm \nabla}\times {\mathbf H}=\partial_t{\mathbf D},\label{max1}\\
&{\bm \nabla}\times {\mathbf E}=-\partial_t{\mathbf B},\label{max2}\\
&{\bm \nabla}{\mathbf D}=0,\label{max3}\\
&{\bm \nabla}{\mathbf B}=0.\label{max4}
\end{align}
\end{subequations}
and the constitutive relations: ${\mathbf D}=\epsilon\epsilon_0 {\mathbf E}$, and ${\mathbf B}=\mu\mu_0 {\mathbf H}$ (with $\epsilon=\mu=1$ in regions $I$ and $III$).
They lead to the standard wave equations: $\Box {\mathbf E}=0$, $\Box {\mathbf B}=0$. Their solutions in the form of plane waves with the assumption, without loss of generality, that the unbounded (guided) propagation takes place in the $x$ direction are well known and recalled below. We use the obvious notation, in which $k_\perp$ denotes the wave-vector component perpendicular to the slab and $k_\parallel$ that along the slab. Since inside the media $\epsilon\mu >1$, it may happen that
\begin{equation}
\omega^2\epsilon\mu-k_\parallel^2=k_\perp^2>0,
\label{din}
\end{equation}
while outside
\begin{equation}
\omega^2-k_\parallel^2=-\alpha^2<0.
\label{dout}
\end{equation}
This situation corresponds to an evanescent wave.

For allowed modes the following standard boundary conditions between electromagnetic fields in the slab and in the surrounding vacuum must be satisfied:
\begin{subequations}
\begin{eqnarray}
{\mathbf D}^{I}_\perp={\mathbf D}^{II}_\perp,\;\;\; {\mathbf D}^{II}_\perp={\mathbf D}^{III}_\perp,\;\;\; {\mathbf E}^{I}_\parallel={\mathbf E}^{II}_\parallel,\;\;\; {\mathbf E}^{II}_\parallel={\mathbf E}^{III}_\parallel,\nonumber\\
\label{bcsde}\\
{\mathbf B}^{I}_\perp={\mathbf B}^{II}_\perp,\;\;\; {\mathbf B}^{II}_\perp={\mathbf B}^{III}_\perp,\;\;\; {\mathbf H}^{I}_\parallel={\mathbf H}^{II}_\parallel,\;\;\; {\mathbf H}^{II}_\parallel={\mathbf H}^{III}_\parallel\nonumber\\
\label{bcsbh}
\end{eqnarray}
\end{subequations}
for $z=\pm d/2$ respectively.

Consider first the $TE^x$ evanescent modes. Due to the mirror symmetry $z\mapsto -z$ these modes can be shown to be either even in $z$ ($+$) or odd ($-$). The electric field has only one component, parallel to the walls of the slab (i.e. in the $y$ direction), so the conditions~(\ref{bcsde}) can easily be exploited, giving~\cite{sc,yeh}
\begin{equation}
\label{TEsyme1}
{\mathbf E}={\mathbf e}_y e^{-i\, \omega t+i k_\parallel x}\left\{\begin{array}{lc}\pm E_0'e^{\alpha z},&$I$,\\
E_0 f_\pm(z),&$II$,\\
E_0'e^{-\alpha z},&$III$,\end{array}\right.
\end{equation}
with
\begin{equation}
E_0'=E_0e^{\alpha d/2}f_\pm(d/2),
\label{TEes1}
\end{equation}
where $f_+(z)=\cos(k_\perp z)$ and $f_-(z)=\sin (k_\perp z)$. The upper signs in all formulas refer to even and lower ones for odd modes. Symbols $I$, $II$, and $III$ are abbreviations for the conditions $z \leq -d/2$, $|z| < d/2$ and $z \geq d/2$ respectively. We do not go into the details, since the wave propagation in dielectrics is well known and worked out by many authors. Our objective is to concentrate rather on the relations between $2D$ and $3D$ description, so we recall these well-known results only to the exent required for this goal.

The corresponding magnetic field may be easily found by acting with the curl operator. We have then
\begin{eqnarray}
&&{\mathbf B}=e^{-i\, \omega t+i k_\parallel x}\nonumber\\
&&\times\left\{\begin{array}{lc}\pm\frac{i\,}{\omega}\, E_0'(\alpha {\mathbf e}_x-i\, k_\parallel {\mathbf e}_z)e^{\alpha z},& $I$,\\
\frac{i\,}{\omega}\, E_0(\mp k_\perp f_\mp(z) {\mathbf e}_x - i\, k_\parallel f_\mp(z){\mathbf e}_z),& $II$,\\
-\frac{i\,}{\omega}\, E_0'(\alpha {\mathbf e}_x+i\, k_\parallel {\mathbf e}_z)  e^{-\alpha z},& $III$,\end{array}\right.\nonumber\\
\label{TEsymm1}
\end{eqnarray}

Using boundary conditions for the magnetic field in~(\ref{TEsymm1}) again the conditions~(\ref{TEes1}) are obtained, as well as additional ones:
\begin{equation}
E_0'=\pm\frac{k_\perp}{\mu\alpha}\,E_0e^{\alpha d/2}f_\mp(d/2).
\label{TEms1}
\end{equation}

The conditions imposed on the wave vectors can be written in the following dimensionless form:
\begin{equation}
\eta=\pm\frac{1}{\mu}\,\xi\tan^{\pm 1}\frac{\xi}{2},
\label{TEcond1}
\end{equation}
where
\begin{equation}
\xi=k_\perp d,\;\; \eta=\alpha d,\;\; \Omega=\omega d
\label{xieta}
\end{equation}

The additional constraint arises from the wave-equation by subtracting~(\ref{din}) and (\ref{dout}):
\begin{equation}
\xi^2+\eta^2=(n^2-1)\Omega^2=:\tau^2,
\label{circle}
\end{equation}
with $\xi,\eta>0$. The intersections of this quarter of a circle with the curves described by~(\ref{TEcond1})  drawn in the $\xi\eta$ plane define the allowed values of $k_\perp$ (for given $d$, $\omega$ and $n$), for which evanescent modes in the slab exist.

Now let us consider the $TM^x$ mode. The magnetic field is now oriented along the $y$ axis. One finds:
\begin{equation}
\label{TMsymm1}
{\mathbf B}_\pm={\mathbf e}_y e^{-i\, \omega t+i k_\parallel x}\left\{\begin{array}{lc}\pm B_0'e^{\alpha z},&$I$,\\
B_0 f_\pm(z),&$II$,\\
B_0'e^{-\alpha z},& $III$,\end{array}\right.
\end{equation}
where
\begin{equation}
B_0'=\frac{1}{\mu}\,B_0e^{\alpha d/2}f_\pm(d/2).
\label{TMbs1}
\end{equation}

The apparent asymmetry between formulas~(\ref{TMbs1}) and~(\ref{TEes1}) disappears, if one uses the ${\mathbf H}$ field instead of ${\mathbf B}$. The final formulas shall be entirely symmetric.

For electric field one obtains
\begin{eqnarray}
&&{\mathbf E}_\pm=e^{-i\, \omega t+i k_\parallel x}\nonumber\\
&&\times\left\{\begin{array}{lc}\mp \frac{i\,}{\omega}\, B_0'(\alpha {\mathbf e}_x-i\, k_\parallel {\mathbf e}_z)e^{\alpha z}&$I$,\\
\frac{i\,}{\omega\epsilon\mu}\, B_0(\pm k_\perp f_\mp(z) {\mathbf e}_x +i\, k_\parallel f_\pm(z){\mathbf e}_z)&$II$,\\
\frac{i\,}{\omega}\, B_0'(\alpha {\mathbf e}_x+i\, k_\parallel {\mathbf e}_z)  e^{-\alpha z}&$III$,\end{array}\right. .\nonumber\\
\label{TMsyme1}
\end{eqnarray}

\begin{figure*}[t!]
\begin{center}
\includegraphics[width= 0.75\textwidth,angle=0]{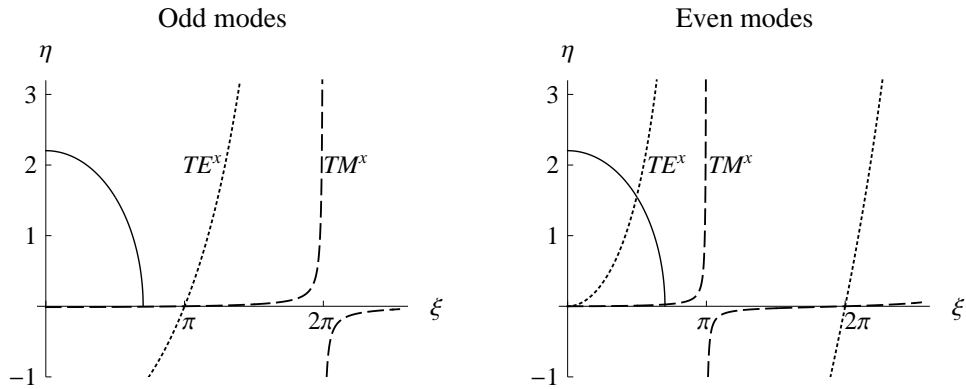}
\end{center}
\caption{Evanescent modes in the dielectric slab with $\epsilon> 1$ and $\mu=1$. Dotted lines correspond to the $TE^x$ and dashed ones to the $TM^x$ modes. The solid line represents the condition~(\ref{circle}).}
\label{slabmodes}
\end{figure*}

Boundary conditions applied to~(\ref{TMsyme1}) give:
\begin{equation}
B_0'=\pm\frac{k_\perp}{\epsilon\mu\alpha}\,B_0e^{\alpha d/2}f_\mp(d/2),
\label{TMms1}\\
\end{equation}
In the variables $\xi,\eta$ we obtain now the following conditions:
\begin{equation}
\eta=\pm\frac{1}{\epsilon}\,\xi\tan^{\pm 1}\frac{\xi}{2}.
\label{TMcond1}
\end{equation}
together with~(\ref{circle}).

In Fig.~\ref{slabmodes} the graphical solutions of the equations for the dielectric slab without any magnetic properties ($\epsilon> 1$, $\mu=1$) are shown. On the first plot the odd modes are visualized. If $\tau<\pi$, then no odd evanescent modes are possible within the slab (both dotted and dashed lines have $\eta<0$ in this region, as seen from~(\ref{TEcond1}) and~(\ref{TMcond1})). This can be understood as follows. For small $\omega$ it is not possible (at the distance of the slab thickness $d$) to fit a slowly varying function to antisymmetric boundary condition. The situation can be improved by increasing the value of the refractive index. Then the wavelength becomes shorter, which would be reflected in the figure by the enlargement of the circle radius. For larger values of $n$ it is easier to catch evanescent modes. The smaller the value of $n$, the closer the direction of the wave vector must be to the boundary. In consequence, $k_\perp$ becomes small, and this means the slow $z$-dependence of fields.

For even modes shown on the second plot, the situation is different. Even if $\tau<\pi$ we have exactly one $TE^x$ and one $TM^x$ mode captured. From the points of intersection we see, however, that the value of $\alpha$ is quite different for these two modes. Large value of $\alpha$ means, that the $TE^x$ mode is well localized within the slab (in the transverse direction). Its extremely small value for the $TM^x$ mode indicates, that it is strongly delocalized and can hardly be be treated as bound inside. This may be advantageous from the point of view of energy loss in a waveguide but cannot be described within the two-dimensional theory. For the data used to plot this graph one sees, that in the case of $TE^x$ the value of $\eta$ is about $1.6$, which means that the damping distance equals approximately $0.6d$. For $TM^x$ this value is two orders of magnitude larger.

The condition $\tau<\pi$ in terms of thickness, refractive index and wavelength in the media ($\lambda_n$) reads
\begin{equation}
d<\frac{\lambda_n}{2}\, \frac{1}{\sqrt{1-1/n^2}}.
\label{condd}
\end{equation}
For a slightly dielectric material, i.e. when $n$ is close to unity or in other words for `weakly guiding media', the right-hand side becomes large and the slab may be thick. This can be understood in terms of geometrical optics, since in such a case it is very difficult to keep the propagating wave inside the slab. The critical angle becomes close to $\pi/2$ and the majority of modes, not sliding along the surface, escape from the media. For large values of $n$, as in Fig.~\ref{slabmodes}, the thickness must be less than a half of the wavelength.

It may seem a bit paradoxical that we treat the high refractive index as unfavorable. After all, it is then easier to keep the waves in the plate. But from the point of view of the accurateness of a two-dimensional model, when $n\gg 1$ these modes become too  numerous. On the other hand the value of $n$ too close to unity leads to small values of $\alpha$ and electromagnetic fields extend to a large distance, which turns out to be essential for curved media. Hence, one could conclude, that the optimal value for $n$ is moderate (in all figures in this work $n=13$).

By choosing the appropriately small value of $\tau$, we can expel all modes except one from the slab (i.e. they cease to be the evanescent ones). This is the case, we are interesting in; exactly one $TE$-wave surface mode (called the fundamental mode) is guided along the slab. There is, however, no lower bound; one always has exactly one $TE^x$ mode captured (apart from the possible various directions along $xy$ plane). This situation most closely corresponds to the two-dimensional propagation.

For small $\tau$ the parameter $\xi$ is small too. This means that $k_\perp z\ll 1$ (for $-d/2<z<d/2$) and the only evanescent mode that remains, has the following approximate form inside the slab:
\begin{eqnarray}
&&{\mathbf E}={\mathbf e}_y E_0e^{-i\, \omega t+i k_\parallel x},\label{redte}\\
&&{\mathbf B}=
{\mathbf e}_z\,\frac{E_0 k_\parallel}{\omega}\, e^{-i\, \omega t+i k_\parallel x}=
{\mathbf e}_zE_0 n e^{-i\, \omega t+i k_\parallel x}.
\nonumber
\end{eqnarray}
It corresponds to the two-dimensional plane wave given by~(\ref{mo1}), but with $k_\parallel=\omega n$ and not $\omega$. However, a certain price has to be paid; taking $\tau\ll 1$, we simultaneously strongly decrease $\alpha$. This means, that the captured mode will have large tails outside the slab.

For a magnetic slab, when $\epsilon=1$ and $\mu> 1$, the $TE^x$ and $TM^x$ modes of Fig.~\ref{slabmodes} interchange their roles. The value of $\alpha$ in the $TM^x$ mode becomes large, and $TE^x$ mode gets delocalized. In the case of small value of $\tau$, for the trapped surface mode $TM^x$ we approximately obtain:
\begin{equation}
{\mathbf B}={\mathbf e}_y B_0e^{-i\, \omega t+i k_\parallel x},\;\; {\mathbf E}=
 -{\mathbf e}_zB_0e^{-i\, \omega t+i k_\parallel x}.
\label{redtm}
\end{equation}
As will be discussed below such waves cannot be obtained within the standard $2D$ electromagnetism; it is incomplete to properly describe a thin layer.

\subsection{A mathematical model: two-dimensional plane}\label{plane}

For a dielectric slab one can use as a model the two-dimensional electromagnetism described by the Lagrange density ${\mathcal L}=\textstyle{\frac{\epsilon_0}{2}}\left({\bm E}^2-B^2\right)$ leading to the reduced Maxwell equations:
\begin{subequations}\label{plme}
\begin{align}
&\nabla_y H=\partial_t D_x,\label{plme1}\\
&\nabla_x H=-\partial_t D_y,\label{plme2}\\
&\nabla_xE_y-\nabla_yE_x=-\partial_t B,\label{plme3}\\
&\nabla_xD_x+\nabla_yD_y=0,\label{plme4}
\end{align}
\end{subequations}
with $D_x=\epsilon_0E_x$, $D_y=\epsilon_0E_y$ and $B=\mu_0H$. The value of $\epsilon_0$ in $2D$, however, is in general different from that in $3D$, and should be determined from the measurement of the Coulomb force in such a hypothetic  world.

The two-dimensional wave equation, together with~(\ref{plme3}) and~(\ref{plme4}), leads to the plane wave with the only possible polarization corresponding to the $TE^x$ mode in the planar waveguide (in $2D$ we naturally put $n=1$):
\begin{eqnarray}
&&{\mathbf E}(x,y,t)={\mathbf e_y} E_0e^{-i\,\omega t+i\, k_\parallel x},\label{mo1}\\
&&B(x,y,t)= \frac{E_0 k_\parallel}{\omega}\,e^{-i\,\omega t+i\, k_\parallel x}=E_0 e^{-i\,\omega t+i\, k_\parallel x}.
\nonumber
\end{eqnarray}
The $TM^x$ mode cannot be modeled in this way. However, as pointed out in Sec.~\ref{rl}, it does not play an important role in a nonmagnetic slab.

The fields ${\mathbf E}$ and $B$ are expressed in the standard way through the three-component electromagnetic potential $[\Phi, {\mathbf A}]$ (however, only one component is independent after gauge fixing and after exploiting the Gauss law):
\begin{eqnarray}
&&E_x=-\partial_x\Phi-\partial_tA_x,\;\; E_y=-\partial_y\Phi-\partial_tA_y,\nonumber\\
&&B=\partial_xA_y-\partial_yA_x,
\label{r1v}
\end{eqnarray}

For the description of a thin magnetic slab we need a theory, in which the magnetic field has two components and it is a vector tangent to the space while the electric field is a scalar. This theory is governed by the following set of Maxwell equations:
\begin{subequations}
\begin{eqnarray}
&&\nabla_xH_y-\nabla_yH_x=\partial_t D,\label{plmm1}\\
&&\nabla_x E=\partial_t B_y,\label{plmm2}\\
&&\nabla_y E=-\partial_t B_x,\label{plmm3}\\
&&\nabla_xB_x+\nabla_yB_y=0,\label{plmm4}
\end{eqnarray}
\end{subequations}
instead of those given by~(\ref{plme1})-(\ref{plme4}). The constitutive equations are $B_x=\mu_0H_x$, $B_y=\mu_0H_y$ and $D=\epsilon_0E$. These fields may be generated from a single scalar potential $A$, via the equations:
\begin{equation}
E=-\partial_t A,\;\; B_x=-\partial_y A,\;\; B_y=\partial_xA.
\label{r2v}
\end{equation}
and the appropriate Lagrange density would then be
\begin{equation}
{\cal L}=\frac{\epsilon_0}{2}\left[(\partial_t A)^2-(\partial_x A)^2-(\partial_y A)^2 \right],
\label{lagr2}
\end{equation}
describing a massless, neutral, scalar field. Introducing sources to generate such waves in this theory, however, would be puzzling, since conserved currents flowing within the surface do not generate waves such as~(\ref{redtm}), but rather~(\ref{redte}), unless one admits the existence of magnetic charges. Some other kind of sources, not respecting the continuity equation in 2D, would be necessary.

Equations~(\ref{plmm1})-(\ref{plmm4}) lead to the following plane waves propagating along the $x$ axis:
\begin{equation}
 E(x,y,t)= -E_0 e^{-i\,\omega t+k_\parallel x},\;\; {\mathbf B}(x,y,t)={\mathbf e_y} B_0e^{-i\,\omega t+k_\parallel x}.
\label{mo2}
\end{equation}
in accordance with~(\ref{redtm}), if one neglects large tails outside the slab.

\section{Hollow fiber}\label{cyl}

\subsection{Propagation in a real cylindrical layer}\label{cl}

\begin{figure}[h!]
\begin{center}
\includegraphics[width= 0.25\textwidth,angle=0]{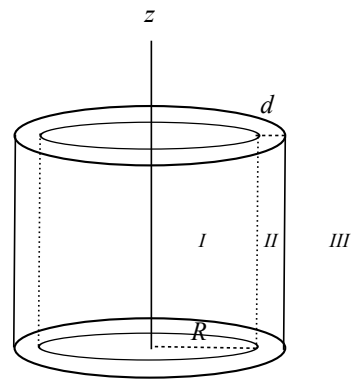}
\end{center}
\caption{Dielectric/magnetic cylindrical layer and definitions of regions $I$, $II$, $III$.}
\label{figcyl}
\end{figure}

In Fig.~\ref{figcyl} we show an electro-magnetic hollow cylinder of the thickness $d$ and internal radius $R$.  We look for evanescent modes trapped in region $II$, depending on the values of $\epsilon$, $\mu$, $\omega$ and $d$.

In the case of a structure with cylindrical symmetry the simplest approach is to describe electromagnetic waves in terms of two vector potentials~\cite{bal}, since Maxwell equations are sourceless. One can define them, writing:
\begin{equation}
{\mathbf B}={\bm \nabla}\times {\mathbf A},\;\; {\mathbf D}={\bm \nabla}\times {\mathbf C}.
\label{2pot}
\end{equation}

The space itself is flat and only the boundary conditions will later introduce the cylindrical topology. Hence one simply has here ${\bm D}=\epsilon\epsilon_0 {\bm E}$ and ${\bm B}=\mu\mu_0 {\bm H}$ although for vector components (not for a vector as a geometric object) these relations in curvilinear coordinates may become more complicated.

From Maxwell equations for monochromatic waves we obtain:
\begin{subequations}
\begin{eqnarray}
{\bm E}=i\,\omega{\bm A}+\frac{i\,}{\epsilon\mu\omega}\,{\bm \nabla}({\bm\nabla}{\bm A})+\frac{1}{\epsilon\epsilon_0}\,{\bm \nabla}\times{\bm C},\label{epot}\\
{\bm B}=-\frac{i\,\omega\mu}{\epsilon_0}{\bm C}-\frac{i\,}{\epsilon\epsilon_0\omega}\,{\bm \nabla}({\bm\nabla}{\bm C})+{\bm \nabla}\times{\bm A},\label{bpot}
\end{eqnarray}
\end{subequations}
where the potentials ${\bm A}$ and ${\bm C}$ satisfy identical Helmholtz equations:
\begin{equation}
\bigtriangleup {\bm A}+\omega^2\epsilon\mu{\bm A}=0,\;\;
\bigtriangleup {\bm C}+\omega^2\epsilon\mu{\bm C}=0.\label{eqAC}
\end{equation}
In what follows, for the description of various modes we will use cylindrical coordinates $\rho$, $\varphi$ and $z$.

\subsection{$TE^z$ modes.}\label{modetez}

For the $TE^z$ modes one can choose
\begin{equation}
{\bm A}=0,\;\; {\bm C}=C_z(\rho,\varphi,z){\bm e}_z=\tilde{C}_z(\rho)e^{i\, m\varphi+i\, k_z z}{\bm e}_z,
\label{tez}
\end{equation}
In region $II$ the coefficient $\tilde{C}_z(\rho)$ satisfies:
\begin{equation}
\left[\partial_\rho^2+\frac{1}{\rho}\,\partial_\rho+\left(\omega^2\epsilon\mu-k_z^2-\frac{m^2}{\rho^2}\right)\right]\tilde{C}_z(\rho)=0.
\label{frade}
\end{equation}
In regions $I$ and $III$ the equation is similar, but with $\epsilon\mu\mapsto 1$. We are looking for evanescent modes, i.e. satisfying~(\ref{din}) and~(\ref{dout}),
where now $k_\parallel=k_z$.

\begin{figure*}[htb!]
\begin{center}
\includegraphics[width=0.75\textwidth,angle=0]{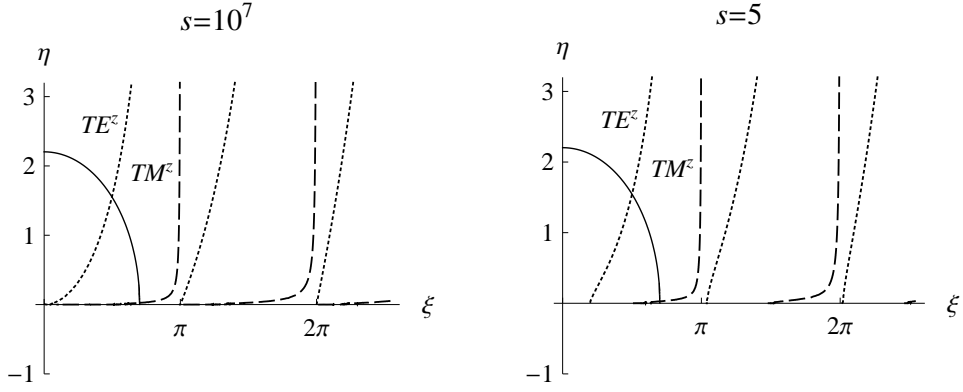}
\end{center}
\caption{Evanescent modes in dielectric hollow cylinder ($\epsilon > 1$ and $\mu=1$) for the parameter $s=10^7$ (first plot) and $s=5$ (second plot).}
\label{cylindermodes}
\end{figure*}

The equation~(\ref{frade}) is the Bessel equation of the order $m$ with the solution:
\begin{equation}
\label{sol1}
\tilde{C}_z(\rho)=\left\{\begin{array}{lc}C_{z0}^{I}I_m(\alpha \rho),& $I$,\\
C_{z0}^{II}J_m(k_\perp \rho)+{C_{z0}'}^{II}Y_m(k_\perp \rho),&$II$,\\
C_{z0}^{III}K_m(\alpha \rho),& $III$,\end{array}\right.
\end{equation}
where those ill-behaved at $\rho=0$ or $\rho\rightarrow\infty$ have been rejected and symbols $I$, $II$, $III$ refer accordingly to $0\leq\rho\leq R$, $R<\rho<R+d$, and $\rho\geq R+d$. The constant coefficients may also be, if needed, complex numbers. Please, note, that if $k_z=0$, the condition~(\ref{dout}) cannot be satisfied. There are no evanescent modes of that kind running in circles. It is a reflection of the nonexistence of photon bound states. This will not be the case in $2D$ propagation.

Out of~(\ref{sol1}) the electromagnetic fields can be generated by simple differentiation. For the electric field one obtains:
\begin{equation}
\label{sole1}
{\bm E}=\left\{\begin{array}{lc}C_{z0}^{I}\Big[\frac{i\, m}{\rho}\,I_m(\alpha \rho){\mathbf e}_\rho-\alpha I_m'(\alpha \rho){\mathbf e}_\varphi\Big],& $I$,\\
\frac{1}{\epsilon}\Big[\frac{i\, m}{\rho}(C_{z0}^{II}J_m(k_\perp \rho)+{C_{z0}'}^{II}Y_m(k_\perp \rho)){\mathbf e}_\rho\\-k_\perp(C_{z0}^{II}J_m'(k_\perp \rho)+{C_{z0}'}^{II}Y_m'(k_\perp \rho)){\mathbf e}_\varphi\Big],
 & $II$,\\
C_{z0}^{III}\Big[\frac{i\, m}{\rho}\,K_m(\alpha \rho){\mathbf e}_\rho-\alpha K_m'(\alpha \rho){\mathbf e}_\varphi\Big],& $III$,\end{array}\right.
\end{equation}
and it is visible that $z$ component does not appear, as we wished. The common factor $e^{-i\, \omega t+i\, m \varphi +i\, k_z z}$ has been omitted. The magnetic field, in turn, has in this mode all three components:
\begin{widetext}
\begin{equation}
\label{solb1}
{\bm B}=\left\{\begin{array}{lc}C_{z0}^{I}\Big[\frac{\alpha k_z}{\omega}\,I_m'(\alpha \rho){\mathbf e}_\rho+\frac{i\, m k_z}{\omega\rho}\,I_m(\alpha \rho){\mathbf e}_\varphi+\frac{i\, \alpha^2}{\omega}\, I_m(\alpha \rho){\mathbf e}_z\Big],& $I$,\\
\frac{1}{\epsilon}\Big[\frac{k_\perp k_z}{\omega}(C_{z0}^{II}J_m'(k_\perp \rho)+{C_{z0}'}^{II}Y_m'(k_\perp \rho)){\mathbf e}_\rho
+\frac{i\, m k_z}{\omega\rho}(C_{z0}^{II}J_m(k_\perp \rho)+{C_{z0}'}^{II}Y_m(k_\perp \rho)){\mathbf e}_\varphi& $II$,\\
-\frac{i\, k_\perp^2}{\omega}(C_{z0}^{II}J_m(k_\perp \rho)+{C_{z0}'}^{II}Y_m(k_\perp \rho)){\mathbf e}_z\Big],\\
 C_{z0}^{III}\Big[\frac{\alpha k_z}{\omega}\,K_m'(\alpha \rho){\mathbf e}_\rho+\frac{i\, m k_z}{\omega\rho}\,K_m(\alpha \rho){\mathbf e}_\varphi+\frac{i\, \alpha^2}{\omega}\, K_m(\alpha \rho){\mathbf e}_z\Big],& $III$.\end{array}\right.
\end{equation}
\end{widetext}

The boundary conditions~(\ref{bcsde}) and~(\ref{bcsbh}) must currently be met for $\rho=R$ and $\rho=R+d$. It is, however, a well-known fact~\cite{carson,chew} that this set of requirements is inconsistent unless we put $m=0$ (it is the {\em depolarization effect}, which leads to mixing $TE^z$ and $TM^z$ waves). We recall this fact here since it does not appear in the two-dimensional model.

Let us now assume that wave propagates along the cylinder ($m=0$) and let us write down the four independent conditions, after having introduced parameters defined by~(\ref{xieta})), as well as $ s=R/d$.  This last parameter will be important, since it allows us to investigate the behavior of the system for differently bent layers. The boundary conditions may be given the following matrix form:
\begin{eqnarray}
\left[\begin{array}{cccc}\eta I_0'&-\frac{1}{\epsilon}\,\xi J_0'& -\frac{1}{\epsilon}\,\xi Y_0'&0\\
\eta^2 I_0&\frac{1}{\epsilon\mu}\,\xi^2 J_0&\frac{1}{\epsilon\mu}\,\xi^2 Y_0&0\\
0&-\frac{1}{\epsilon}\,\xi \tilde{J}_0'&-\frac{1}{\epsilon}\,\xi \tilde{Y}_0'&\eta K_0'\\
0&\frac{1}{\epsilon\mu}\,\xi^2 \tilde{J}_0&\frac{1}{\epsilon\mu}\,\xi^2 \tilde{Y}_0&\eta^2 K_0\end{array}\right]\,\left[\begin{array}{c} C_{z0}^{I}\\ C_{z0}^{II}\\ {C_{z0}'}^{II}\\ C_{z0}^{III}\end{array}\right]=0,\nonumber\\
\label{bctez}
\end{eqnarray}
where we denoted
\begin{eqnarray}
&&I_m=I_m(\eta s),\;\;\;\; I_m'=I_m'(\eta s),\nonumber\\
&&K_m=K_m(\eta (s+1)),\;\;\;\; K_m'=K_m'(\eta (s+1)),\nonumber\\
&&J_m=J_m(\xi s),\;\;\;\; Y_m=Y_m(\xi s),\nonumber\\
&&\tilde{J}_m=J_m(\xi (s+1)),\;\;\;\; \tilde{Y}_m=Y_m(\xi (s+1)).
\label{not}
\end{eqnarray}

The condition for the existence of solutions is:
\begin{equation}
\det M_{TE}=0,
\label{mte}
\end{equation}
where $M_{TE}$ is the matrix in~(\ref{bctez}). The allowed evanescent modes correspond to intersections (in the $\xi\eta$ plane) of the curve~(\ref{mte}) represented in Fig.~\ref{cylindermodes} with dotted line with that of~(\ref{circle}), drawn in solid, which retains its previous form.

\subsection{$TM^z$ modes.}\label{modetmz}

For the $TM^z$ modes we choose
\begin{equation}
{\bm A}=A_z(\rho,\varphi,z){\bm e}_z=\tilde{A}_z(\rho)e^{i\, m\varphi+i\, k_z z}{\bm e}_z,\;\; {\bm C}=0,
\label{tmz}
\end{equation}
Due to the symmetry between electric and magnetic fields the solution can be obtained by merely introducing to the former one the necessary changes. The magnetic field has now the form almost identical to~(\ref{sole1}) with the obvious replacement of constants $C_{z0}^{I}, C_{z0}^{II}, {C_{z0}'}^{II}, C_{z0}^{III}$ by $A_{z0}^{I}, A_{z0}^{II}, {A_{z0}'}^{II}, A_{z0}^{III}$ and with the identification:
\begin{equation}
{\mathbf E}^{I}\mapsto {\mathbf B}^{I},\;\; \epsilon{\mathbf E}^{II}\mapsto {\mathbf B}^{II},\;\; {\mathbf E}^{III}\mapsto {\mathbf B}^{III}.
\label{subEB}
\end{equation}
Similarly for electric field we can use~(\ref{sole1}) with the following substitutions:
\begin{equation}
-{\mathbf B}^{I}\mapsto {\mathbf E}^{I},\;\; -\frac{1}{\mu}{\mathbf B}^{II}\mapsto {\mathbf E}^{II},\;\; -{\mathbf B}^{III}\mapsto {\mathbf E}^{III}.
\label{subBE}
\end{equation}

As a result the matrix $M_{TM}$ differs from $M_{TE}$ only by the interchange $\epsilon\leftrightarrow\mu$. Curves corresponding to the condition
\begin{equation}
\det M_{TM}=0,
\label{mtm}
\end{equation}
are drawn in Fig.~\ref{cylindermodes} with dashed lines. On the first plot the parameter $s$ has been fixed as $10^7$, which corresponds to a large cylinder. We plot modes for dielectric media only for $\epsilon> 1$ and $\mu=1$, since for the magnetic layer (when $\epsilon=1$ and $\mu> 1$) the graphs become identical after the interchange of dashed and dotted lines. In this latter case, the captured wave does not have its $2D$ counterpart, and requires going beyond the standard two-dimensional electrodynamics, as discussed in the following subsection.

When the value of the parameter $\tau$ is properly adjusted, we have exactly one $TE^z$ and $TM^z$ evanescent mode. As it was true for the slab, the values of $\alpha$ differ significantly  for these two modes. $\epsilon> 1$ and $\mu=1$ leads to the large value of $\alpha$ for $TE^z$ mode and the wave is  localized within the cylindrical shell and for the $TM^z$ mode this value is small and it becomes delocalized.

The conclusions for $s\gg 1$, are very similar to those for the dielectric slab. The system of propagating modes exactly corresponds to that shown in Fig.~\ref{slabmodes}. This is not surprising, since a large cylinder locally can be treated as a slab. This can directly be observed after putting $\rho=R+\zeta$ in~(\ref{sole1}) and in (\ref{solb1}) and also after expanding the expressions for $\zeta/R\ll 1$, apart from oscillatory factors.

The modes visualized in Fig.~\ref{cylindermodes} are apparently twice as many as those in~\ref{slabmodes}, but it is so because now all of them are drawn: corresponding to odd and even modes of the plate. Of course we consider here only the modes with azimuthal number $m=0$, which describe the propagation along the $x$ axis in the slab. Other possibilities are considered below.

For a tight fiber ($s=5$) the situation changes. There exists a limiting value of $\tau$ (i.e. a certain $\omega_{\mathrm min}$), below which no evanescent modes arise. This is clearly visible on the second plot of Fig.~\ref{cylindermodes}, where the first dotted line starts at certain $\xi>0$. This effect is similar to the existence of low frequency cutoff in cylindrical waveguide~\cite{collin}. Its emergence depends only on the value of $s$ or, in other words, on the surface curvature. For a large and thin cylinder this effect is negligible. 

\begin{figure}[t!]
\begin{center}
\includegraphics[width=0.4\textwidth,angle=0]{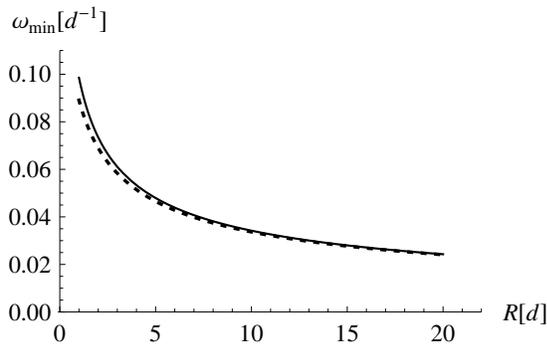}
\end{center}
\caption{The minimal value of $\omega$ in units of $d^{-1}$ of the $TE^z$ evanescent mode, as a function of the cylinder surface radius $R$ in units of $d$. The dotted line represents the approximate result given by the formula~(\ref{os}). In $2D$ this value is constantly zero.} \label{omegamin} 
\end{figure}

In Fig.~\ref{omegamin} the dependence of $\omega_{\mathrm min}(R)$ is drawn. The threshold for the wave propagation is raised with the increasing value of the curvature. In turn, for large $R$, $\omega_{\mathrm min}$ gradually decreases to zero, which is the value appropriate for the slab. It should be pointed out that for the $2D$ surface mode no such threshold (for this mode) arises and $\omega_{\mathrm min}$ is permanently equal to $0$.

Using asymptotic values for the Bessel functions appearing
in~(\ref{bctez}) one can find the approximate value of $\omega_{\mathrm min}$:
\begin{equation}
\omega_{\mathrm min}(R)\approx \frac{2}{d\sqrt{(n^2-1)(1+2R/d)}}.
\label{os}
\end{equation}

This result almost perfectly agrees with the exact graph shown in Fig.~\ref{omegamin}. The line corresponding to~(\ref{os}) is then drawn as dotted.
The shift of the position of zeros to the right, visible in Fig.~\ref{cylindermodes}, decreases for more excited modes, i.e. those more strongly oscillating in the perpendicular direction (large $k_\perp$). In this case even for relatively small value of $s$ the allowed mods better correspond to those in the slab.

For small values of $k_\perp \rho$ the electromagnetic fields for $TE^z$ evanescent mode inside the layer may be approximately written as:
\begin{eqnarray}
&&{\mathbf E}=-{\mathbf e}_\varphi E_0e^{-i\, \omega t+i k_\parallel x},\label{redtez}\\
&&{\mathbf B}=
{\mathbf e}_\rho \frac{E_0 k_\parallel}{\omega}\,e^{-i\, \omega t+i k_\parallel x}=
{\mathbf e}_\rho E_0 ne^{-i\, \omega t+i k_\parallel x}
\nonumber
\end{eqnarray}
where $E_0=2{C_{z0}'}^{II}/(\pi\epsilon R)$, and for $TM^z$:
\begin{eqnarray}
&&{\mathbf B}={\mathbf e}_\varphi B_0e^{-i\, \omega t+i k_\parallel x},\label{redtmz}\\
&&{\mathbf E}=
{\mathbf e}_\rho \frac{B_0 k_\parallel}{n\omega}\,e^{-i\, \omega t+i k_\parallel x}=
{\mathbf e}_\rho B_0 e^{-i\, \omega t+i k_\parallel x}
\nonumber
\end{eqnarray}
with $B_0=-2\epsilon {A_{z0}'}^{II}/(\pi\epsilon R)$.

\begin{figure*}[t!]
\begin{center}
\includegraphics[width= 0.75\textwidth,angle=0]{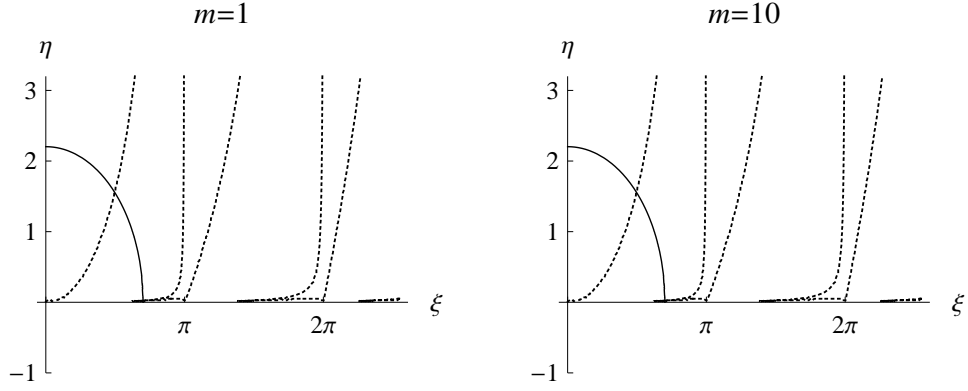}
\end{center}
\caption{Hybrid evanescent modes for large $R$ or thin layer ($s=10^7$). }
\label{cylh}
\end{figure*}

\subsection{Hybrid modes}\label{hybrid}

In order to account for the `skewed' propagation and to satisfy the cylindrical boundary conditions with $m\neq 0$ we need to consider a more complicated mode called a {\em hybrid} mode~\cite{chew}. When a light ray moves along a helical path, the polarization changes due to the interaction with the walls and we cannot have a solution with one pure polarization. This limitation does not appear in two-dimensional propagation.

A hybrid mode $H^z$ is a combination of $TE^z$ and $TM^z$ modes:
\begin{subequations}\label{hyb}
\begin{align}
{\bm A}&=A_z(\rho,\varphi,z){\bm e}_z=\tilde{A}_z(\rho)e^{i\, m\varphi+i\, k_z z}{\bm e}_z,\label{hybA}\\
 {\bm C}&=C_z(\rho,\varphi,z){\bm e}_z=\tilde{C}_z(\rho)e^{i\, m\varphi+i\, k_z z}{\bm e}_z,\label{hybC}
\end{align}
\end{subequations}
and electric and magnetic fields are sums of those given earlier. There are now $8$ constants to be determined and $12$ equations resulting from boundary conditions: $3$ for $\mathbf E$ and $3$ for $\mathbf B$ on the inner surface, and similarly on the outer one. Since $4$ conditions for perpendicular components result from the tangential ones, which are their linear combinations, we are left with $8$ equations and the corresponding matrix $M_H$ is now $8\times 8$. If we denote $\gamma=m(1+\eta^2/\Omega^2)^{1/2}$, this matrix may be written in the form:
\begin{widetext}
\begin{eqnarray}
\left[\begin{array}{cccccccc}\eta I_m'&-\frac{\xi}{\epsilon} J_m'&-\frac{\xi}{\epsilon}\, Y_m'&0&-\frac{\gamma}{s}\,I_m& \frac{\gamma}{\epsilon\mu s}\,J_m&\frac{\gamma}{\epsilon\mu s}\,Y_m&0\\
0&-\frac{\xi}{\epsilon} \tilde{J}_m'&-\frac{\xi}{\epsilon} \tilde{Y}_m'&\eta
K_m'&0&\! \frac{\gamma}{\epsilon\mu (s+1)}\,\tilde{J}_m&\!\frac{\gamma}{\epsilon\mu (s+1)}\,\tilde{Y}_m&\!-\frac{\gamma}{s+1}\,K_m\\
\frac{\gamma}{s}\, I_m&-\frac{\gamma}{\epsilon\mu s}\,J_m&-\frac{\gamma}{\epsilon\mu s}\,Y_m&0&-\eta I_m'&\frac{\xi}{\mu} J_m'&\frac{\xi}{\mu}\, Y_m'&0\\
0&\!-\frac{\gamma}{\epsilon\mu (s+1)}\,\tilde{J}_m&\!-\frac{\gamma}{\epsilon\mu (s+1)}\,\tilde{Y}_m&\frac{\gamma}{s+1}\,K_m&0&\frac{\xi}{\mu}\, \tilde{J}_m'&\frac{\xi}{\mu}\, \tilde{Y}_m'&-\eta K'_m\\
0&0&0&0&\eta^2 I_m&\frac{\xi^2}{\epsilon\mu}\,J_m&\frac{\xi^2}{\epsilon\mu}\,Y_m&0\\
0&0&0&0&0&\frac{\xi^2}{\epsilon\mu}\,\tilde{J}_m&\frac{\xi^2}{\epsilon\mu}\,\tilde{Y}_m&\eta^2 K_m\\
\eta^2 I_m&\frac{\xi^2}{\epsilon\mu}\,J_m&\frac{\xi^2}{\epsilon\mu}\,Y_m&0&0&0&0&0\\
0&\frac{\xi^2}{\epsilon\mu}\,\tilde{J}_m&\frac{\xi^2}{\epsilon\mu}\,\tilde{Y}_m&\eta^2 K_m&0&0&0&0\end{array}\right].\nonumber\\
\label{hyma}
\end{eqnarray}
\end{widetext}
It acts on the vector of coefficients:
\begin{equation}
\left[\begin{array}{cccccccc}C_{z0}^{I}& C_{z0}^{II}&{C_{z0}'}^{II}&C_{z0}^{III}&A_{z0}^{I}&A_{z0}^{II}&{A_{z0}'}^{II}& A_{z0}^{III}\end{array}\right],
\label{vca}
\end{equation}
where apart from~(\ref{not}) we defined:

\begin{eqnarray}
&&J_m'=J_m'(\xi s),\;\; Y_m'=Y_m'(\xi s),\;\; \tilde{J}_m'=J_m'(\xi (s+1)),\nonumber\\
&&\tilde{Y}_m'=Y_m'(\xi (s+1)).\nonumber\\
\label{yy}
\end{eqnarray}

The condition $\det M_H=0$, for a large cylinder ($s=10^7$), leads to the curves shown in Fig.~\ref{cylh}. They are drawn for a dielectric layer with $\epsilon > 1$ and $\mu=1$ and for the two values of azimuthal quantum number: $m=1$ and $m=10$ turning out to be insensitive to the value of $m$. The propagation with $m\neq 0$ corresponds just to the propagation in the slab in a direction other than $x$ and there is no qualitative difference between the waves with various $m$. The situation will change, when $R$ is not large as compared to $d$, and the cylindrical layer should not be treated as being locally a plane due to its large curvature.

\begin{figure*}[htb]
\begin{center}
\includegraphics[width= 0.75\textwidth,angle=0]{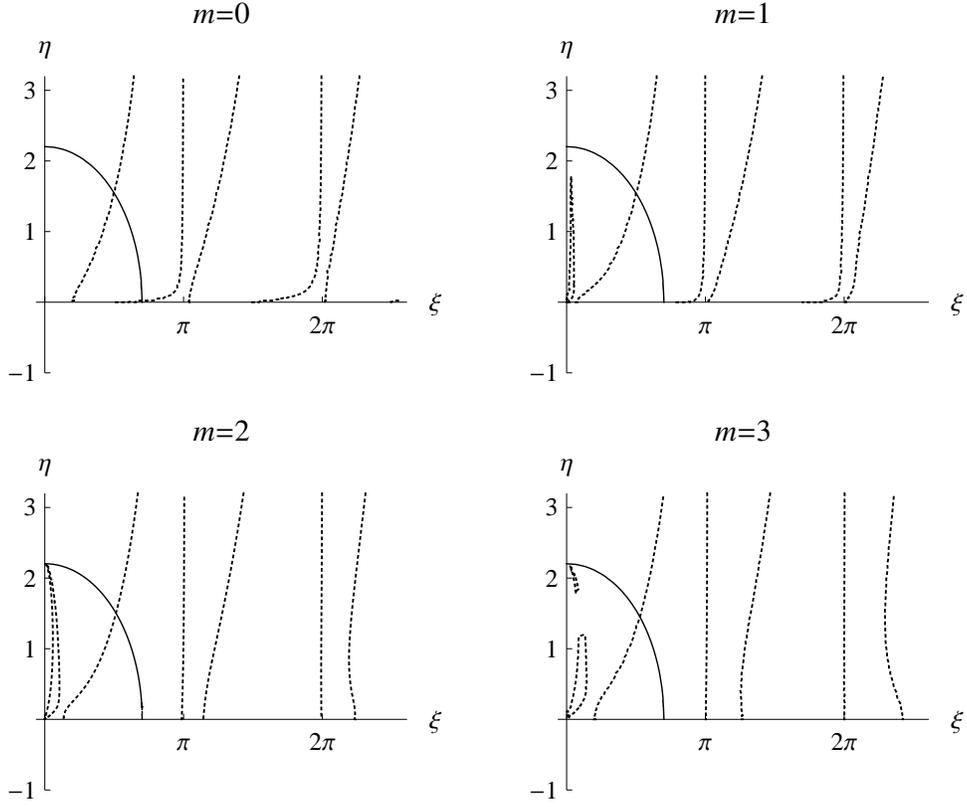}
\end{center}
\caption{Hybrid evanescent modes for small $R$ ($s=5$) for the increasing values of $m$. The closed loops near the origin never cross the solid line and are not significant.}
\label{cylh1}
\end{figure*}

In Fig.~\ref{cylh1} we show the evanescent modes for a tight ($s=5$) dielectric surface. The subsequent plots are drawn for the increasing values of the azimuthal quantum number. For the propagating modes, according to~(\ref{frade}), we must have:
\begin{equation}
\omega^2n^2-\frac{m^2}{R^2}>0.
\label{cp}
\end{equation}
which in terms of dimensionless variables reads:
\begin{equation}
m<\Omega s n
\label{cp1}
\end{equation}
and corresponds roughly to the condition that $m$ be smaller than the number of wavelengths $\lambda_n$ fitting in the perimeter. The same requirement (with refractive index $n=1$) appears naturally in the $2D$ surface model. Apart from it we again observe the shift of all curves to the right, a cut-off phenomenon known from the cylindrical rod~\cite{collin}. This cutoff seems to disappear for $m=1$. This is reminiscent of the fact that for a rod the cut-off is nonexistent for the so called $HE_{11}$ mode.

The additional curves in the form of closed loops never cross the line~(\ref{circle}) and therefore do not correspond to any mode. In fact they can be eliminated if one exploits~(\ref{circle}) in the equation $\det M_H=0$, before drawing the latter.

\begin{figure*}[htb!]
\begin{center}
\includegraphics[width= 0.75\textwidth,angle=0]{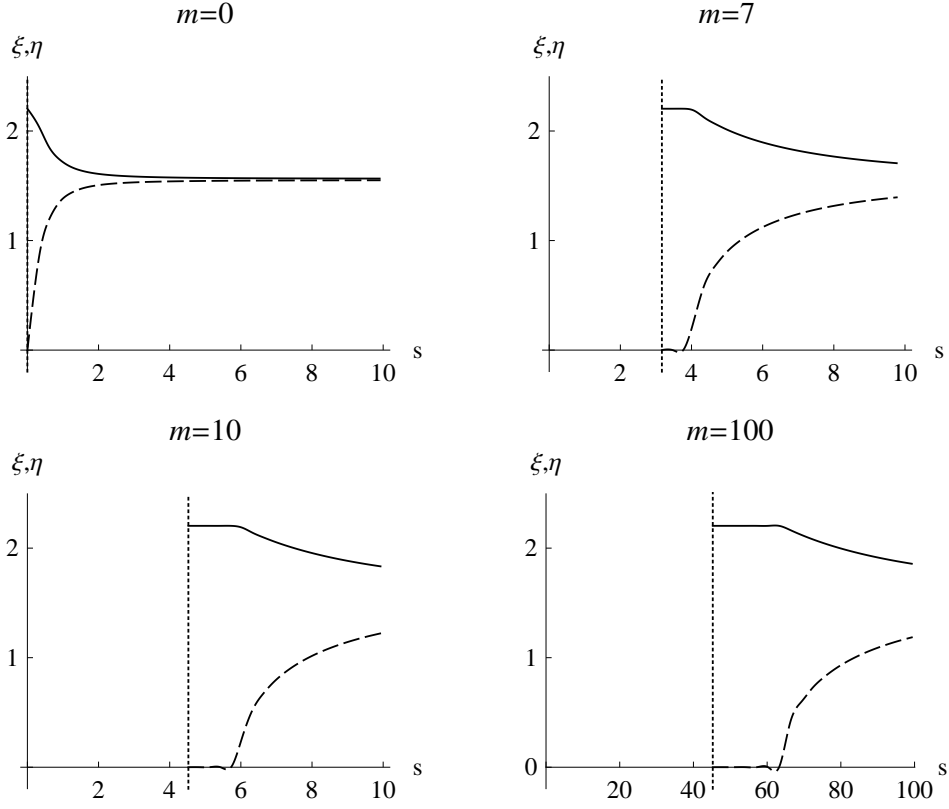}
\end{center}
\caption{The dependence of the allowed values of $\xi$ (solid line) and $\eta$ (dashed line) on $s$ for different values of $m$. The dotted line represents the threshold imposed by~(\ref{cp1}).}
\label{cylhm}
\end{figure*}

In Fig.~\ref{cylhm} the dependence of the allowed values of $\xi$ and $\eta$ (i.e. the crossing points for the dotted and solid curves of Fig.~\ref{cylh1}) for various values of the azimuthal number $m$ are drawn as functions of $s$. The vertical dotted lines correspond to the condition~(\ref{cp1}), which must be satisfied in both cases (if the wave is to propagate): in the $2D$ world and in the layer (considering the $2D$ world with the same light velocity). It should be noted, that in the physical layer, for large values of $m$ the parameter $\eta$ becomes zero even above this threshold (and so does $\alpha$), so the wave becomes radiative and escapes. The threshold is then shifted up. This effect corresponds to the one observed in Fig.~\ref{cylindermodes} and~\ref{omegamin} for $TE^z$ mode. It may then not be possible to create evanescent modes of large $m$ in a tight dielectric cylindrical layer even if the two-dimensional model admits them. For large $m$ the wave resembles a wave running in circles, which was excluded in the equation~(\ref{frade}).

\subsection{A mathematical model: two-dimensional cylindrical surface}\label{cylinder}

In cylindrical coordinates $\varphi$, $z$ and $\rho=R$, the reduced metric tensor has the form:
\begin{equation}
\hat{g}_{2D}=\left[\begin{array}{cc} R^2 & 0 \\ 0 & 1 \end{array}\right].
\label{gcyl}
\end{equation}
and the fields are connected by the following constitutive relations:
\begin{equation}
D_\varphi=\epsilon_0 \frac{1}{R}\, E_\varphi,\;\; D_z=\epsilon_0 R E_z,\;\; B=\mu_0 R H.
\label{debhc}
\end{equation}

The Maxwell equations now read:
\begin{subequations}\label{clme}
\begin{align}
&\nabla_z H=\partial_t D_\varphi,\label{clme1}\\
&\nabla_\varphi H=-\partial_t D_z,\label{clme2}\\
&\nabla_\varphi E_z-\nabla_z E_\varphi=-\partial_t B,\label{clme3}\\
&\nabla_\varphi D_\varphi+\nabla_zD_z=0,\label{clme4}
\end{align}
\end{subequations}
Manipulating them one can easily get the wave-equation for the magnetic field $B$:
\begin{equation}
\partial_t^2B=\frac{1}{R^2}\,\partial_\varphi^2B+\partial_z^2B
\label{wavcylb}
\end{equation}
and similarly for the components of electric field. The simplest monochromatic solution has the form ($m=0,1,2,\ldots$):
\begin{subequations}
\begin{align}
&B(\varphi,z,t)=B_0e^{-i\,\omega t+i\, m \varphi+i\, k_\parallel z},\label{bsolc}\\
&E_\varphi(\varphi,z,t)=-\frac{k_\parallel B_0}{\omega}\,e^{-i\,\omega t+i\, m \varphi+i\, k_\parallel z},\label{e1solc}\\
&E_z(\varphi,z,t)=\frac{m B_0}{\omega R^2}\,e^{-i\,\omega t+i\, m \varphi+i\, k_\parallel z},\label{e2solc}
\end{align}
\end{subequations}
where
\begin{equation}
\omega^2=\frac{m^2}{R^2}+k_\parallel^2.\label{omk}
\end{equation}
This means, that for a fixed value of $\omega$, $m$ cannot be too large or $R$ cannot be too small ($R>m/\omega$) for a propagating wave. This wave corresponds to that found in the layer, explicitly given in~(\ref{redtez}) for $m=0$. Contrary to the propagation in the layer, any $m$ satisfying~(\ref{omk}) is admitted in~$2D$. One should also note that the mode describing a wave running in circles is not excluded. If $k_\parallel=0$ the condition~(\ref{omk}) simply means that the circumference of the cylinder comprises exactly $m$ wavelengths.

This theory, however, is not sufficient to describe the evanescent mode in the magnetic layer, given by~(\ref{redtmz}). Instead of using~(\ref{clme1})-(\ref{clme4}) and~(\ref{debhc}), we have to exploit the following set of equations:
\begin{subequations}\label{cclme}
\begin{align}
&\nabla_\varphi H_z-\nabla_z H_\varphi=\partial_t D,\label{cclme1}\\
&\nabla_\varphi E=-\partial_t B_z,\label{cclme2}\\
&\nabla_z E=\partial_t B_\varphi,\label{cclme3}\\
&\nabla_\varphi B_\varphi+\nabla_zB_z=0,\label{cclme4}
\end{align}
\end{subequations}
together with:
\begin{equation}
D=\epsilon_0 R E,\;\; B_\varphi=\mu_0 \frac{1}{R}\, H_\varphi,\;\; B_z=\mu_0 R H_z.
\label{debhc2}
\end{equation}
They lead to a monochromatic solution:
\begin{subequations}
\begin{align}
&B_\varphi(\varphi,z,t)=\frac{B_0k_\parallel}{\omega}\,e^{-i\,\omega t+i\, m \varphi+i\, k_\parallel z},\label{b1solc}\\
&B_z(\varphi,z,t)=-\frac{m  B_0}{\omega}\,e^{-i\,\omega t+i\, m \varphi+i\, k_\parallel z},\label{b2solc}\\
&E(\varphi,z,t)=B_0e^{-i\,\omega t+i\, m \varphi+i\, k_\parallel z},\label{esolc}
\end{align}
\end{subequations}
to be compared with~(\ref{redtmz}).

At the end one should note that any form of the two-dimensional electromagnetism, neither defined by~(\ref{clme}) nor~(\ref{cclme}) is sufficient to represent correctly the hybrid modes.

\section{Summary}\label{conc}
In the present paper we dealt with the propagation of electromagnetic waves in thin dielectric and magnetic layers. For the dielectric slab, by adjusting the parameters, we can expel all the modes except exactly one $TE^z$ and one $TM^z$ evanescent mode from the media. However the latter has large, slowly decaying tails outside the slab. For the magnetic layer the role and behavior of these two modes are interchanged. Our main goal was to establish if the propagation may be described within two-dimensional electrodynamics. We have found that it roughly corresponds to it, although the complete description requires now two alternative versions of the $2D$ electromagnetism. In the first, traditional,  version the electric field has a vector character and lives within the surface, and magnetic field is a scalar. In the second version the magnetic field must have two components, and electric field has a scalar character. This important aspect has commonly been neglected.

In the case of a curved surface, in addition to the effects described above, some new ones emerge, which put additional limits on the applicability of the $2D$ description. 
\begin{enumerate}
\item Even for the simplest $TE$ mode traveling along the cylinder there appears a certain limiting value of $\omega_{\mathrm min}$, below which no evanescent modes are possible. 
\item The value of this threshold increases with the layer curvature. For a large cylinder radius $\omega_{\mathrm min}$ decays as $R^{-1/2}$. No such cut-off appears in the cylindrical surface, no matter how tight the cylinder is.
\item For helical propagation, in the case of a layer, the only possible modes are the so called `hybrid' ones. No purely $TE$ or $TM$ modes are admitted, although in the $2D$ surface they freely travel. 
\item Contrary, the hybrid modes can be represented in two dimensions in none of the alternative forms of electromagnetism.
\item Even the mode running in circles (forbidden in the layer) is allowed in $2D$ unless the wavelength fits integer number of times into the perimeter.
\item In the layer the threshold effects connected with the curvature shift the $2D$ value of $\omega_{\mathrm min}$ from $m/R$ to larger values. This shift increases with raising azimuthal number $m$, except at $m=1$.
\end{enumerate}

These results and limitations can be of certain importance for the propagation of electromagnetic waves within thin layers of nontrivial geometry, particularly of large curvature, which does not allow for explicit mathematical solutions. In such circumstances the two-dimensional description remains the main option.

\section*{Acknowledgments}
The authors would like to thank to Professor Iwo Bialynicki-Birula for the inspiration, many elucidating discussions, careful reading of the paper and suggested improvements.
The work was supported by the Polish National Science Center Grant No. 2012/07/B/ST1/03347.

\end{document}